\documentclass[twocolumn,preprintnumbers,amsmath,amssymb,showpacs]{revtex4}

\pdfoutput=1 % for arXiv

\usepackage{graphicx}% Include figure files
\usepackage{bm}% bold math
\usepackage{dcolumn}
\usepackage{amssymb}
\usepackage{amsmath}
\usepackage{amsfonts}
\usepackage{color}
\usepackage{epsfig}
\begin{document}
% -------- Title -----------
\title{
     Magnetic Instability on the Surface of Topological Insulators}
\author{
    Yuval Baum and Ady Stern }
\affiliation{
     Department of Condensed Matter Physics, Weizmann Institute of Science, Rehovot 76100, Israel}

% -------- Abstract -----------
\begin{abstract}
Gapless surface states that are protected by time reversal symmetry and charge conservation are among the manifestations of $3D$ topological insulators.
In this work we study how electron-electron interaction may lead to spontaneous breaking of time reversal symmetry on surfaces of such insulators. We find that
a critical interaction strength exists, above which the surface is unstable to spontaneous formation of magnetization, and study the dependence of this critical interaction strength on temperature and chemical potential. 
\end{abstract}
\maketitle

%\section{INTRODUCTION}%-------------------------------------------
Topological states of matter are one of the main themes of modern condensed matter physics.
A recent important  discovery in this field are Topological Insulators \cite{Hassan,Zhang}.
In three dimensions Topological Insulators are  electronic materials that have a bulk gap like an ordinary insulator, but have conducting states on their surface \cite{Hassan,mele}. The gapless spectrum on the surface arises from the nontrivial topological order of the band structure \cite{FKM,moore}. It is composed of an odd number of Dirac cones in the Brillouin zone of the two dimensional surface. These Dirac cones cannot be gapped for as long as time reversal symmetry and charge conservation are maintained.

In this work we explore the conditions under which the surface of a strong topological insulator may lower its energy by a spontaneous breaking of time reversal symmetry through the formation of a uniform spin polarization. The way energy may be saved by spin polarization  is most easily understood for the case where there is one Dirac cone on each surface, and the chemical potential lies at the Dirac point. For such a case, two sources for energy gain may be identified: first, if the spin polarization introduces a mass term to the Dirac Hamiltonian, the filled electronic states are lowered in energy, while the energy of the unfilled states is raised. And second - as usual for electrons - if all spins are polarized, two electrons cannot reside at the same point, and the short-distance energy cost is reduced. This is not the case in the absence of magnetization. Even though the half filled Dirac cone includes just one electron for every momentum state, the variation of the spin's direction with that of the momentum leads to a non-zero probability of two electrons being at the same point, and thus to a gain of interaction exchange energy by spin polarization. 

In our analysis, carried out within the Hartree-Fock approximation, we find that there is a critical interaction strength above which the spin polarized ground state is favorable in energy compared to the non-interacting ground state. We find this to be the case for every value of the chemical potential, for all temperatures much smaller than the bulk energy gap, and for both contact interaction and Coulomb interaction. Of the possible spin polarizations we find that the lowest energy corresponds to an out-of-plane polarization, which introduces an energy gap to the Dirac cone at the Dirac point. For contact interaction, we find the phase transition from the unpolarized to the polarized phase to be of second order both as a function of interaction strength and as a function of temperature. Finally, we find the energy gap to rise sharply as temperature is lowered below the temperature $T_c$ of the phase transition, and reach values comparable to $T_c$ at about $T\approx 0.9 T_c$. We note that a gapped Dirac cone was recently observed  by angle resolved photo emission (ARPES) in the topological insulator phase \cite{Ando}, but the information available is not sufficient to judge whether this observation is related to the mechanism we study.

We start with the non-interacting Hamiltonian. ARPES measurements carried out on two topological insulators, Bi$_2$Te$_3$ and Bi$_2$Se$_3$, have found that while near the Dirac point the spectrum is Dirac-like, an hexagonal warping of the Fermi surface occurs away from the Dirac point \cite{Chen,Hsieh}.
This phenomenon was modeled by  Fu \cite{Fu}, who suggested an effective Bloch Hamiltonian of the surface including a cubic term:
\begin{equation}\label{freeH}
\mathcal{H}(\boldsymbol{k})=v_0(k_x\sigma_y-k_y\sigma_x)+\lambda k^3cos(3\theta)\sigma_z
\end{equation}
where $v_0$ is the electron velocity near the Dirac point, $\lambda$ is the warping parameter and $\sigma$ are the Pauli matrices.
The above spectrum contains two branches that meet at the Dirac point.
This Hamiltonian is time-reversal symmetric. At the time-reversal invariant point $k=0$ the spectrum is degenerate and the degeneracy is protected by the time-reversal symmetry.

The interaction part of the Hamiltonian is
%\section{Mean-Field Approach}%---------------------------------------------------------------------------------------------
\begin{align}\label{Eq:intH}
\mathcal{H_{I}}=\int{d\boldsymbol{r}d\boldsymbol{r'}\psi_{\sigma}^\dag{\scriptstyle (\boldsymbol{r})}\psi_{\sigma}{\scriptstyle (\boldsymbol{r})}V{\scriptstyle (\boldsymbol{r}-\boldsymbol{r'})}\psi_{\sigma'}^\dag{\scriptstyle (\boldsymbol{r'})}\psi_{\sigma'}{\scriptstyle (\boldsymbol{r'})}}
\end{align}
Here $\psi_{\sigma}^\dag (\boldsymbol{r})$ is the creation operator of an electron at point $\boldsymbol{r}$ and spin $\sigma$.
We will assume that the following operator has a non vanishing expectation value:
\begin{align}\label{Eq:expval}
\langle \psi_{\sigma}^\dag{\scriptstyle (\boldsymbol{r})}\psi_{\sigma'}{\scriptstyle (\boldsymbol{r'})}\rangle\equiv M_{\sigma,\sigma'}{\scriptstyle (\boldsymbol{r}-\boldsymbol{r'})}
\end{align}
 Strictly speaking,  we can interpret $M$ as the magnetization only when $\boldsymbol{r}=\boldsymbol{r'}$. However, the expectation value (\ref{Eq:expval}) breaks time reversal symmetry even when $\boldsymbol{r}\ne\boldsymbol{r'}$. We confine ourselves to uniform states, in which $M$ will depend only on
$\boldsymbol{R} \equiv \boldsymbol{r}-\boldsymbol{r'}$. This case is most relevant when the Fermi-energy is located at the vicinity of the Dirac point, otherwise spin-wave formation is an interesting possibility.

The expectation value (\ref{Eq:expval}) may be written in a matrix form as
$M_{\sigma,\sigma'}{\scriptstyle (\boldsymbol{r}-\boldsymbol{r'})}=M_{0}{\scriptstyle (\boldsymbol{R})}\cdot I+\vec{M}{\scriptstyle (\boldsymbol{R})}\cdot\vec{\sigma}$, but since we are interested only in the spin part of the interaction and in particular the out-of-plane spin ($z$ direction), we will assume that only $M_z\ne0$, hence:
$M_{\sigma,\sigma'}{\scriptstyle (\boldsymbol{r}-\boldsymbol{r'})}\equiv M{\scriptstyle (\boldsymbol{R})}\sigma_z$. In-plane magnetization will be discussed in a later part of the paper.

By denoting
$F_{\boldsymbol{k}}=\int{d\boldsymbol{R} V{\scriptstyle (\boldsymbol{R})}M{\scriptstyle (\boldsymbol{R})}\exp{(-i\boldsymbol k\cdot\boldsymbol R)}}$, we can write the mean-field Hamiltonian as:
\begin{align} \label{Eq:mfH}
H^{MF}\equiv \int{d\boldsymbol{R} V{\scriptstyle (\boldsymbol{R})}M^2{\scriptstyle (\boldsymbol{R})}}+\int{\frac{d\boldsymbol k}{(2\pi)^2}c^\dag_{\boldsymbol k}\hat{h}(\boldsymbol k)c_{\boldsymbol k}}
\end{align}
where
\begin{align}\label{Eq:h}
\hat{h}=v_0(k_x\sigma_y-k_y\sigma_x)+\big(\lambda k^3\cos3\theta-2F_{\boldsymbol{k}}\big)\sigma_z \nonumber
\end{align}

For a self-consistent  determination of  the out-of-plane magnetization, we denote the eigenvalues and the corresponding eigenstates of $\hat{h}$ as $\epsilon_\pm(\boldsymbol k)$ and $\left(\begin{array}{c}
\chi_{\uparrow}, \chi_{\downarrow}
\end{array}\right)_\pm$ respectively. The $\pm$ signs refer to the positive and negative energy states.
Hence, the magnetization per unit volume is:
\begin{equation}\label{Eq:symmag}
m=\frac{1}{\Omega}\sum_{\boldsymbol k}(\langle\sigma_{z-}\rangle f_{\scriptstyle{\epsilon_{-}}}+\langle\sigma_{z+}\rangle f_{\scriptstyle{\epsilon_{+}}})
\end{equation}
where
\begin{align}
\langle\sigma_{z\pm}\rangle=
|\chi_{\uparrow}|_\pm^2-|\chi_{\downarrow}|_\pm^2 \nonumber
\end{align}
and $f_{\scriptstyle{\epsilon}}$ is the Fermi-Dirac function at energy $\epsilon$.
The trivial solution ($m=0$)  always exists, but we are interested in non trivial solutions of this equation.
The total energy per unit volume is:
\begin{align}\label{Eq:symE}
E_m=\int{d\boldsymbol{R} V{\scriptstyle (\boldsymbol{R})}M^2{\scriptstyle (\boldsymbol{R})}}+\frac{1}{\Omega}\sum_{\boldsymbol k}(\epsilon_-f_{\scriptstyle{\epsilon_{-}}}+\epsilon_+f_{\scriptstyle{\epsilon_{+}}})
\end{align}
We will define the energy cost to have finite magnetization as $\Delta E=E_m-E_0$.
If we find a non trivial solution to Eq.~(\ref{Eq:symmag}) that minimizes $E_m$ (gives $\Delta E<0$), we will conclude that the system forms spontaneous surface magnetization.

Although it is hard to determine the exact validity conditions for this variational procedure and to estimate its accuracy, it is still possible to define a necessary condition.
Since the procedure above relies on the existence of a bulk gap, it is clear that it requires the interaction energy to be smaller than the bulk gap.
We will consider the condition $U_c<\Delta_{bulk}$ as necessary for the above procedure (where $U_c$ is the critical interaction energy and $\Delta_{bulk}$ is the bulk energy gap).
%\section{Contact Interaction ($T=0$)}%-------------------------------------------------

It is simplest to analyze a contact interaction, $V{\scriptstyle (\boldsymbol{R})}=\frac{g}{2}\delta{\scriptstyle (\boldsymbol{R})}$.
At zero temperature Eq.~(\ref{Eq:symmag}) for the magnetization $M{\scriptstyle (\boldsymbol{0})}\equiv m$ becomes:
\begin{equation} \label{Eq:contmag}
m=\int_{k_{F}}\frac{d\boldsymbol k}{(2\pi)^2}\frac{gm-\lambda k^3\cos3\theta}{\sqrt{v_0^2k^2+\big(gm-\lambda k^3\cos3\theta\big)^2}}
\end{equation}
where $d\boldsymbol k=kdkd\theta$ and $k_{\mbox {\scriptsize F}}$ is the Fermi-momentum, which is the solution to $v_0k_{\mbox {\scriptsize F}}+\lambda k_{\mbox {\scriptsize F}}^3=\mu$, with $\mu$ the chemical potential.
In general the integration has an upper cutoff in momentum, $k_\Lambda$, which corresponds to the momentum at which the surface states merge with the three dimensional bands. For $\lambda=0$ all results depend linearly on this cutoff, but for $\lambda\neq0$ we can identify a momentum scale $q_0\equiv \big(\frac{v_0}{\lambda}\big)^{1/2}$  above which the integrand decays as $k^{-3}$. The cutoff dependence then disappears for $k_\Lambda>>q_0$.

We find that there is a critical interaction strength $g_c$ at which a second order phase transition occurs, such that for $g>g_c$ there is an $m\ne0$ solution to (\ref{Eq:contmag}) with $\Delta E<0$. The critical interaction strength is

\begin{equation}
g_c=\frac{\sqrt{\lambda v_0}}{Y(p_\Lambda)-Y(p_{F})}=\frac{g_c(\mu=0)}{1-\frac{Y(p_{F})}{Y(p_\Lambda)}} \label{Eq:gc}
\end{equation}
where $p_{\Lambda,F}=k_{\Lambda,F}\sqrt{\frac{\lambda}{v_0}}$ is a dimensionless parameter and the function $Y$ is defined by:
\begin{equation}
Y(p)=\int_0^{p}\frac{p'd p'}{(2\pi)^2}\int_0^{2\pi}d\theta \frac{p'^2}{\Big(p'^2+\big(p'^3\cos3\theta\big)^2\Big)^{3/2}}
\end{equation}
The function $Y(p_{\Lambda})$ is a continuous, positive and monotonically increasing function of $p_\Lambda$ with the following asymptotic behavior:
\begin{equation} \label{Eq:Y_fit}
Y(p_\Lambda)\sim \left\{
\begin{array}{cl}
\frac{p_\Lambda}{2\pi}  &    ,p_\Lambda<<1\\
0.247 & ,p_\Lambda>>1
\end{array} \right.
\end{equation}

\begin{figure}[b]
\begin{center}
\includegraphics[width=6.5cm,angle=0]{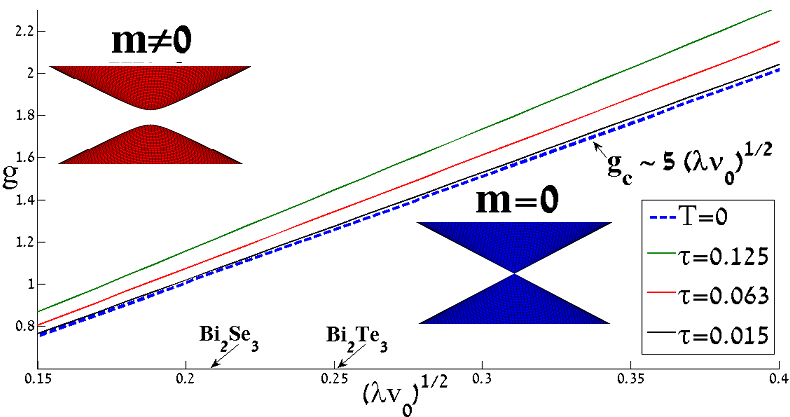}
\caption{ \label{Fig:ins_vs_T_g} %
Phase diagram ($p_\Lambda=2$) for contact interaction at zero (dashed line) and finite temperatures, where $\tau\equiv\frac{T}{E^*}$.
A critical interaction exists, beyond which magnetization formation occurs.
The critical interaction increases as the temperature increases. }
\end{center}
\end{figure}

For $\mu=0$ and for $\lambda\to 0$ Eq.~(\ref{Eq:gc}) becomes $g_c= 2\pi\cdot\frac{v_0}{k_\Lambda}$, while for $k_\Lambda>>({\frac{v_0}{\lambda}})^{1/2}$ Eq.~(\ref{Eq:gc}) becomes $g_c\approx 4.1\cdot\sqrt{v_0\lambda}$. The resulting phase diagram ($\mu=0$) in the large cutoff limit is presented in Fig.~(\ref{Fig:ins_vs_T_g}).

Clearly, for all values of $\lambda$ and $k_\Lambda>k_F$, the critical interaction strength $g_c$ increases as $|\mu|$ increases. In the limit $\lambda \to 0$ Eq.~(\ref{Eq:gc}) gets a simple form:
\begin{equation}
g_c(\mu)=g_c(\mu=0)\cdot \frac{1}{1-\frac{k_F}{k_\Lambda}} \nonumber
\end{equation}
As expected for $k_{\mbox {\scriptsize F}} \to k_\Lambda$ (full band) magnetization formation is not possible. We note, however, that as long as the chemical potential is far from the bulk bands its effect on $g_c$ is rather weak.

For $g\sim g_c$ we can estimate the interaction energy per particle by noticing that $\langle\mathcal{H_{I}}\rangle_{\mbox {\scriptsize per-particle}}=\frac{gn_e}{8}$, where $n_e$ is the electron density on the surface. We will use ARPES data of Bi$_2$Te$_3$ and Bi$_2$Se$_3$ surfaces \cite{BiSe,Hsieh} to compare the critical interaction energy with the bulk energy gap for a typical surface density (table ~\ref{table:1}).
In both materials $U_c<\Delta_{bulk}$, which puts the mean field results within the necessary condition for the validity of the approximation.

\begin{table}[t]
\caption{  Bulk gap vs. $U_c$  for $n_e=5\cdot10^{13}cm^{-2}$ }
\centering
\begin{tabular}{c c c}
\hline\hline
& Bi$_2$Te$_3$ & Bi$_2$Se$_3$ \\ [0.5ex]
\hline
$\Delta_{bulk} $ & 0.16 eV & 0.35 eV \\
$U_c $ & 0.078 eV & 0.066 eV \\ [1ex]
\hline
\end{tabular}
\label{table:1}
\end{table}
%\section{Contact Interaction in Finite Temperature}%-----------------------------------------------------------------------
At non-zero temperatures the transition happens when the free energy of the magnetized surface is lower than that of the unmagnetized one.

It is useful to note that: $\langle\sigma_{z-}\rangle=-\langle\sigma_{z+}\rangle\equiv\langle\sigma_z\rangle$.
For $\mu=0$, the equation for $m$ becomes:
\begin{equation} \label{Eq:magcontT}
m=\int\frac{d\boldsymbol k}{(2\pi)^2} \frac{x(m,{\bf k})\cdot \tanh{\frac{\sqrt{v_0^2k^2+ x_m^2}}{2T}}}{\sqrt{v_0^2k^2+ x_m^2}}
\end{equation}
where $T$ is the temperature and $x(m,{\bf k})=gm-\lambda k^3\cos3\theta$.
The free energy difference ($\Delta f$) per unit volume is
\begin{align}\label{Eq:freeE}
T\int\frac{d\boldsymbol k}{(2\pi)^2}\log\Big(\frac{1+\cosh\big(\sqrt{v_0^2k^2+x_m^2}/T\big)}{1+\cosh\big(\sqrt{v_0^2k^2+(\lambda k^3\cos3\theta)^2}/T\big)}\Big)
%\nonumber
\end{align}
It is natural to define the dimensionless temperature scale $\tau\equiv\frac{T}{E^*}$ where $E^*=\big(\frac{v_0^3}{\lambda}\big)^{1/2}$ is the characteristic energy  scale introduced by hexagonal warping. The resulting phase diagram is presented in Fig.~(\ref{Fig:ins_vs_T_g}).

As expected, the critical interaction strength $g_c$ increases with increasing temperature. We note however that the effect on $g_c$ is rather mild. Again, the phase transition is of second order, both for a fixed temperature as a function of $g$, and for a fixed $g$ as a function of temperature.
%\section{Plane Magnetization}%------------------------------------------------------------------------

A similar analysis may be carried out to examine the conditions for in-plane magnetization to be formed. To that end, we assume a magnetization in the $x-y$ plane, solve the self consistent equation that determines its magnitude and find the conditions under which it leads to an energy gain. As we now show, for typical parameters ($v_0,\lambda,k_\Lambda$)
the critical interaction strength to have in-plane magnetization $g_{c, x-y}$ is smaller than $g_{c,z}$, the critical interaction strength to have a magnetization perpendicular to the plane. We carry out the calculation at zero temperature. We note that an in-plane magnetization does not create a gap in the spectrum. It does however break time-reversal symmetry and hence removes the topological protection of the Dirac-point.

The Bloch Hamiltonian is:
\begin{equation}
\mathcal{H}(\boldsymbol{k})=v_0(k_x\sigma_y-k_y\sigma_x)+\lambda k^3\cos3\theta\sigma_z-g\vec{m}\cdot\vec{\sigma}
\end{equation}
We denote $(m_x,m_y)\equiv\boldsymbol{m_\parallel}$.
In the symmetry broken phase, the system will choose a specific direction. Although the system is not rotationally invariant we find that the angle dependence is very small, and we choose $\boldsymbol{m_\parallel}=m_x$.
The self consistent equation for $m_x$ is then given by:
\begin{equation} \label{Eq:mp}
m_x=\int_{k_F}^{k_\Lambda}\frac{d\boldsymbol k}{(2\pi)^2}\langle\sigma_x(\boldsymbol k)\rangle
\end{equation}
where
\begin{equation}
\langle\sigma_x\rangle=\frac{m_xg+v_0 k \sin{\theta}}{\sqrt{v_0^2k^2+(gm_x)^2+2gv_0 m_x k\sin\theta+(\lambda k^3\cos{3\theta})^2}} \nonumber
\end{equation}

\begin{figure}[b]
\begin{center}
\includegraphics[width=6.5cm,angle=0]{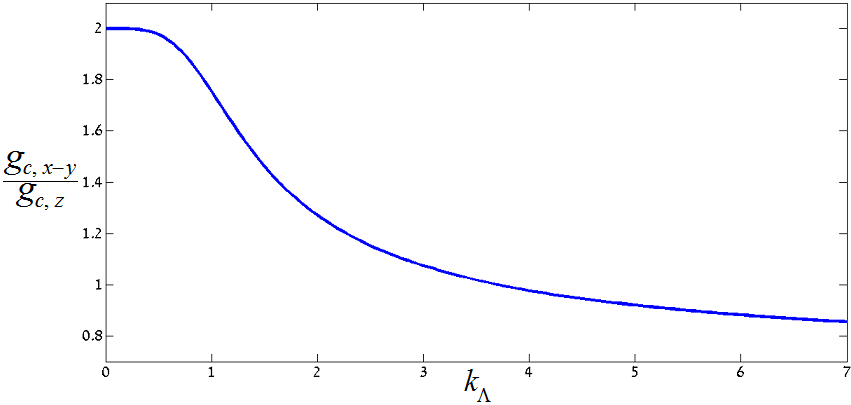}
\caption{ \label{Fig:g_c} %
The ratio between the critical interactions to have in-plane/out-of-plane magnetization as a function of the ratio of $k_\Lambda$ to the warping wave vector scale $(\frac{v_0}{\lambda})^{1/2}$. For typical values the instability to $m_z$ occurs before the one to $m_x$. }
\end{center}
\end{figure}

All that is left now is to solve Eq.~(\ref{Eq:mp}), minimize the energy and deduce the critical interaction $g_{c, x-y}$.
The resulting $g_{c, x-y}$ has the same form as Eq. ~(\ref{Eq:gc}), with $Y(p_\Lambda)$ being replaced by $\Psi(p_\Lambda)$,
where $\Psi(p_\Lambda)$ is a continuous, positive and monotonically increasing function of $p_\Lambda$ with the following asymptotic behavior:
\begin{equation}
\Psi(p_\Lambda)\sim\left\{
\begin{array}{cl}
\frac{p_\Lambda}{4\pi}& ,p_\Lambda<<1\\
\\
 0.352 & ,p_\Lambda>>1
\end{array} \right.
\end{equation}
\newline
The ratio $\frac{g_{c,x-y}}{g_{c,z}}=\frac{Y(p_\Lambda)}{\Psi(p_\Lambda)}$ is presented in Fig.~(\ref{Fig:g_c}) for $\mu=0$.
For $k_\Lambda<3.7 \sqrt{\frac{v_0}{\lambda}}$, the instability to $m_z$ appears before the instability  to  in-plane magnetization. Since this value of $k_\Lambda$
corresponds a warping parameter that is about a hundred times larger than that observed in Bi$_2$Te$_3$, we conclude that the interesting quantity to examine is the magnetization in the z direction. This conclusion does not change when the chemical potential is away from the Dirac point.

Our analysis has so far been focused on contact interaction between the electrons, and has uncovered an instability of the surface to the formation of magnetization, with the instability being strongest for the formation of out-of-plane magnetization at zero temperature and zero chemical potential. The dependence of the critical interaction strength on temperature and chemical potential is however rather weak.
%\section{Coulomb Interaction ($T=0$)}%-----------------------------------------------------------------------------

Motivated by these observations we turn to the long range Coulomb interaction, $V{\scriptstyle (\boldsymbol{R})}=\frac{e^2}{R}$, and study
the instability to the formation of out-of-plane magnetization at zero temperature and zero chemical potential \cite{Peres}.
We assume that $\langle\psi_{\sigma}^\dag{\scriptstyle (\boldsymbol{r})}\psi_{\sigma'}{\scriptstyle (\boldsymbol{r'})}\rangle\equiv M{\scriptstyle (\boldsymbol{R})}\sigma_z$ is an exponentially decaying function of $R$ with a decaying length $\alpha^{-1}$.
This assumption yields two self consistent equations for the magnetization and for $\alpha$:
\begin{align}
m=\int\frac{d\boldsymbol k}{(2\pi)^2}\frac{\frac{4\pi me^2}{\sqrt{k^2+\alpha^2}}-\lambda k^3\cos3\theta}{\sqrt{v_0^2k^2+\big(\frac{4\pi me^2}{\sqrt{k^2+\alpha^2}}-\lambda k^3\cos3\theta\big)^2}}
\end{align}
\begin{align}
\frac{m}{\alpha}=\int\frac{d\boldsymbol k}{(2\pi)^2}
 \frac{\frac{2\pi me^2}{k\sqrt{k^2+\alpha^2}}-\lambda k^2\cos3\theta}{\sqrt{v_0^2k^2+\big(\frac{2\pi me^2}{\sqrt{k^2+\alpha^2}}-\lambda k^3\cos3\theta\big)^2}}
\end{align}

\begin{figure}[t]
\begin{center}
\includegraphics[width=6.5cm,angle=0]{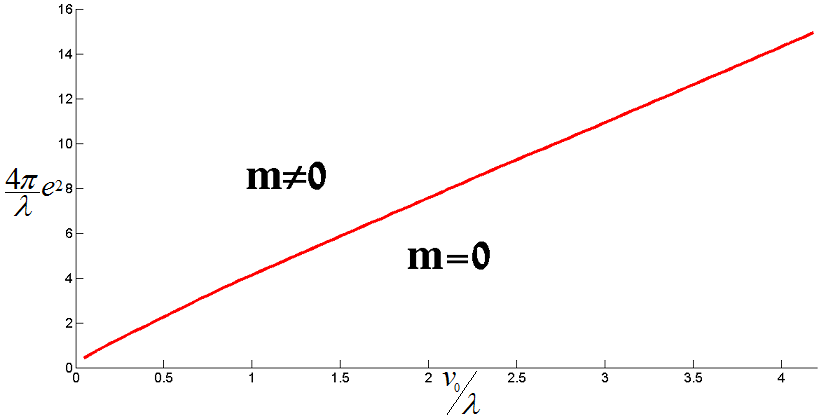}
\caption{ \label{Fig:long2} %
Phase diagram for Coulomb interaction at zero temperature. A critical interaction exists, beyond which magnetization formation occurs.}
\end{center}
\end{figure}

The trivial solution, in which $m=0$ and $\alpha$ is arbitrary, is always an exact solution to the equations above.
After solving numerically for $m$ and $\alpha$ and minimizing the energy, we identify the regions of parameter space in which a non-trivial solution exists, as shown in Fig.~(\ref{Fig:long2}). The values we obtain for $\alpha^{-1}$ range between $1$ and $4$ lattice constants, hence justifying the interpretation of $m$ as the magnetization.

For the Coulomb interaction
the mean interaction energy per particle is of the order of $e^2r_0^{-1}$, where $r_0$ is the mean distance between electrons on the surface. We can estimate $r_0\sim n_e^{-\frac{1}{2}}$.
For a given density we can parameterize the critical interaction strength by defining a critical interaction energy  $U_c=e_c^2r_0^{-1}\sim e_c^2 n_e^{\frac{1}{2}}$. Again, for our approximation to be valid we must have $U_c<\Delta_{bulk}$. We find this condition to be satisfied for both Bi$_2$Te$_3$  and Bi$_2$Se$_3$: Using Fig.~\ref{Fig:long2} and extracting the density from ARPES data we evaluated the critical interaction strength, $e_c^2\sim 0.057e_0^2$ ($0.072e_0^2$) for Bi$_2$Te$_3$ (Bi$_2$Se$_3$), where $e_0$ is the electron charge in vacuum. We find the actual Coulomb interaction on the surfaces of these two materials to be too weak for an instability. The dielectric constant on the two surfaces is close to $40$, as compared to a value of about $15$ for which our approximation leads to an instability.

%\section{Summary and Conclusions}%--------------------------------------
To summarize, we examined here the possibility that time-reversal symmetry is spontaneously broken on the surface of a three dimensional strong topological insulator due to interactions between surface electrons.
We assumed that the surface can be described by a $2D$ effective Dirac Hamiltonian, and treated interactions within the Hartree-Fock approximation. We found that for a strong enough interaction, both of the contact and the Coulomb types, the surface is unstable to the formation of spontaneous magnetization, with the strongest instability being to the formation of magnetization in the direction perpendicular to the surface. For the contact interaction, we found the transition from the unmagnetized to the magnetized surface to be of second order, both at zero temperature as a function of interaction strength and for a fixed interaction strength as a function of temperature. The dependence of the critical interaction strength on temperature and chemical potential, at least for the contact interaction, is rather mild. We found the interaction strength in the two most studied strong topological insulators to be too weak for an instability, but not by a large factor, making the instability an issue that may be relevant for other topological insulators.

The authors thank the US-Israel Binational Science Foundation, the Minerva foundation and Microsoft's station Q for financial support.

\end{document}